\documentclass[runningheads]{llncs}
\usepackage[T1]{fontenc}
\usepackage{amsmath}
\usepackage{graphicx}

\usepackage[T1]{fontenc}
\usepackage{amsmath}
\usepackage{graphicx}

\usepackage{marvosym}
\begin{document}
\title{NEUBORN: The Neurodevelopmental Evolution framework Using BiOmechanical RemodelliNg}
\titlerunning{Neurodevelopmental Evolution Using Biomechanical Remodelling}
%
\authorrunning{N. Baena et al.}

\author{Nashira Baena\inst{1}\textsuperscript{(\Letter)} \and
Mariana da Silva \inst{1} \and
Irina Grigorescu\inst{2} \and
Aakash Saboo\inst{1} \and
Saga Masui\inst{1} \and
Jaques-Donald Tournier\inst{1,2} \and
Emma C. Robinson\inst{1,2}}

\institute{Research Department of Biomedical Computing, School of Biomedical Engineering \& Imaging Sciences, King’s College London \\ \email{paloma.rodriguez\_baena@kcl.ac.uk} 
\and Research Department of Early Life Imaging, School of Biomedical Engineering \& Imaging Sciences, King’s College London
}

%
%
\maketitle              
\begin{abstract}
Understanding individual cortical development is essential for identifying deviations linked to neurodevelopmental disorders. However, current normative modelling frameworks struggle to capture fine-scale anatomical details due to their reliance on modelling data within a population-average reference space. Here, we present a novel framework for learning individual growth trajectories from biomechanically constrained, longitudinal, diffeomorphic image registration, implemented via a hierarchical network architecture. Trained on neonatal MRI data from the Developing Human Connectome Project, the method improves the biological plausibility of warps, generating growth trajectories that better follow population-level trends while generating smoother warps, with fewer negative Jacobians, relative to state-of-the-art baselines. The resulting subject-specific deformations provide interpretable, biologically grounded mappings of development. This framework opens new possibilities for predictive modeling of brain maturation and early identification of malformations of cortical development.

\keywords{Biomechanical Registration \and Cortical Development Analysis \and Cortical Pattern Variability}
\end{abstract}
\section{Introduction}

Population-level comparisons of neurodevelopmental trajectories offer insights into the spatiotemporal evolution of the brain, facilitate the construction of normative growth models \cite{Meng_Li_Wang_Lin_Gilmore_Shen_2018,Duan_Xia_Rekik_Meng_Wu_Wang_Lin_Gilmore_Shen_Li_2019}, and  aid the identification of neurodevelopmental pathologies. Unfortunately, precision mapping of these processes is challenging due to the significant inter-individual variability of secondary and tertiary folds of the cortical surface \cite{Guo_Suliman_Williams_Besenczi_Liang_Dahan_Kyriakopoulou_McAlonan_Hammers_OMuircheartaigh_et_al_2025}.

The cortical folding process is incompletely understood. While primary folds are largely genetically determined and consistent across individuals, the formation of secondary and tertiary folds is believed to be strongly influenced by biomechanical forces. Under this framework, even subtle geometric or mechanical disturbances—such as regional differences in cortical thickness, growth rates, or material properties—can significantly influence the resulting morphology \cite{tallinen_growth_2016}. This mechanical sensitivity gives rise to high variability, such that even genetically identical twins show distinct folding patterns \cite{Bartley_Jones_Weinberger_1997}.

Errors at any stage of cortical folding can result in Malformations of Cortical Development (MCDs), which are associated with neurodevelopmental disorders including intellectual disability, autism spectrum conditions, cerebral palsy, and epilepsy \cite{severino_definitions_2020}. Current approaches for detecting early imaging biomarkers rely on normative models that compare
cortical features within a population-average space  \cite{Meng_Li_Wang_Lin_Gilmore_Shen_2018,Duan_Xia_Rekik_Meng_Wu_Wang_Lin_Gilmore_Shen_Li_2019}, which assumes that all data can be perfectly aligned using highly-regularised, diffeomorphic transformations. Individual variability in cortical folding, however, means that this is generally not the case; for instance, where cortical areas may be represented by two folds in some brains but one fold in others, optimisation of cortical matching becomes ill-posed, and current regularisation frameworks are not designed to handle the extreme deformations that would be required to address this. As a result, more variable, fine-scale gyral and sulcal features are often poorly aligned, resulting in spatial blurring. Ultimately, this reduces the sensitivity of classic normative frameworks to detect subtle, subject-specific biomarkers \cite{Zöllei_Iglesias_Ou_Grant_Fischl_2020,Schmitt_Raznahan_Liu_Neale_2020}. Precision modelling of individualised normative developmental trajectories is therefore essential if we are to truly understand healthy development and, in turn, identify early signs of disease as deviations from these expected trajectories. 

\subsubsection{Related works}
Finite Element Methods (FEM) simulate neurodevelopment from biomechanical models, which assume greater expansion of the cortex relative to sub-cortical tissues below – modelling tissues as hyperelastic materials, which, under tensile stresses, buckle and fold \cite{zarzor_two-field_2021,tallinen_gyrification_2014,budday_mechanical_2017}. Such models can simulate the effects of mechanical perturbations within simple 2D or 3D simulations, to show how pathological alterations during development can lead to malformations in cortical folding \cite{budday_mechanical_2014-1}. However, FEM methods are computationally costly, and depend on incomplete biomechanical models of folding for optimisation, as a result, they can produce fold-like structures but fail to fully capture the complexity of cortical folding observed in individual human brains.

A more direct way to quantify cortical growth is to measure it through longitudinal alignment of MRI scans acquired from the same brain at different developmental stages. However, most current image registration frameworks \cite{Dalca2019UnsupervisedSurfaces,balakrishnan_voxelmorph_2019,avants2009advanced,jenkinson_fsl_2012} are designed with population analyses in mind. This has very different constraints relative to perinatal growth, since alignment between subjects must account for heterogeneity of the cortical shape, while alignment over time involves a much greater expansion of the cortex relative to the tissue of white matter below.

\subsubsection{Contribution}
We propose to build from \cite{Simpson_Cardoso_Modat_Cash_Woolrich_Andersson_Schnabel_Ourselin_2015}, by integrating spatially-adaptive regularisation terms into a learning-based image registration framework. We achieve this by taking inspiration from Da Silva et al. \cite{DaSilva2021DistinguishingNetworks} which proposes a learning-based biomechanical model of brain atrophy. While their approach is limited to small deformations, the framework is general and can be extended to neurodevelopment through appropriate adaption of the biomechanical model~\cite{tallinen_growth_2016}. In this paper, we present our full-brain Neurodevelopmental Evolution framework Using BiOmechnical
RemodelliNg (NEUBORN) and validate it for longitudinal image registration, demonstrating that it can handle large deformations while accurately capturing individual cortical growth trajectories.

\section{Methods}
\subsection{Data, Acquisition \& Preprocessing}
This study leverages longitudinal MRI of 92 preterm neonates scanned shortly after birth and again at term-equivalent age, acquired as part of the Developing Human Connectome Project. Imaging was performed using a 3T Philips Achieva scanner equipped with a 32-channel neonatal head coil. T2-weighted images were acquired using a fast spin-echo sequence with TR/TE = 12,000/156 ms, an in-plane resolution of 0.8 mm × 0.8 mm, a slice thickness of 1.6 mm, and a 0.8 mm overlap. The scans underwent motion correction and reconstruction before bias correction, brain extraction, and Draw-EM tissue segmentation (following the standard dHCP structural pipeline \cite{makropoulosdeveloping2018,makropoulos2014automatic,makropoulos2016regional}). Finally, volumes were linearly aligned using FSLs Linear Image Registration Tool (FLIRT) \cite{jenkinson2001global}.

\begin{figure}
\includegraphics[width=\textwidth]{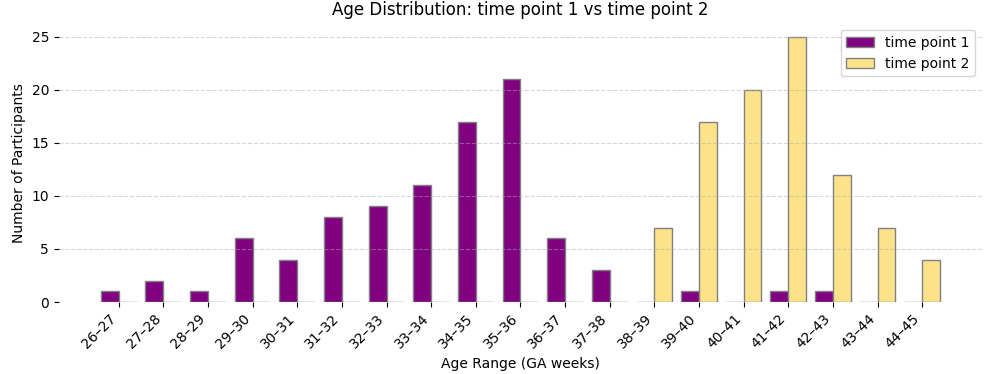}
\caption{Distribution of scan ages for all 92 subjects: In magenta the first scan taken shortly after birth, in yellow the follow-up scan taken at around full term equivalent age.} 
\label{fig3}
\end{figure}

\subsection{Model overview}

We aim to learn a data-driven, biophysically constrained mapping function, $h_\zeta(M, R) = \phi$, where $h$ is a neural network parameterised by $\zeta$, and $\phi$ is the deformation field that warps the moving image $M$ into alignment with the reference, $R$. The network is trained by jointly minimising an image similarity loss between $M \circ \phi$ and $R$, and a biomechanical loss that encourages physiologically plausible deformations.

\subsubsection{Diffeomorphic Deformation Integration}

We model the deformation field $\phi$ as a smooth, invertible mapping that preserves topology (consistent with prior work~\cite{Mok_Chung_2020,Dalca2019UnsupervisedSurfaces,balakrishnan_voxelmorph_2019}). Specifically, $\phi$ is defined as the solution to an Ordinary Differential Equation (ODE) in the form of a Stationary Velocity Field (SVF), $v$, ~\cite{Ashburner_2007}:
\begin{equation}
\frac{d\phi_t(x)}{dt} = v(\phi_t(x)), \quad \phi_0(x) = x
\end{equation}
Integrating this ODE over $t \in [0,1]$ yields a diffeomorphic transformation $\phi_1(x)$. Since direct integration is computationally expensive, we approximate this by scaling the velocity field by $1/2^T$ for a sufficiently large $T $, and recursively composing the resulting small deformations:
\begin{equation}
\phi^{(0)}(x) = x + \frac{v(x)}{2^T}, \quad \phi^{(t+1)}(x) = \phi^{(t)} \circ \phi^{(t)}(x) \quad \text{for } t = 0, \dots, T-1
\end{equation}

\subsubsection{Hierarchical Architecture}
\begin{figure}
\includegraphics[width=\textwidth]{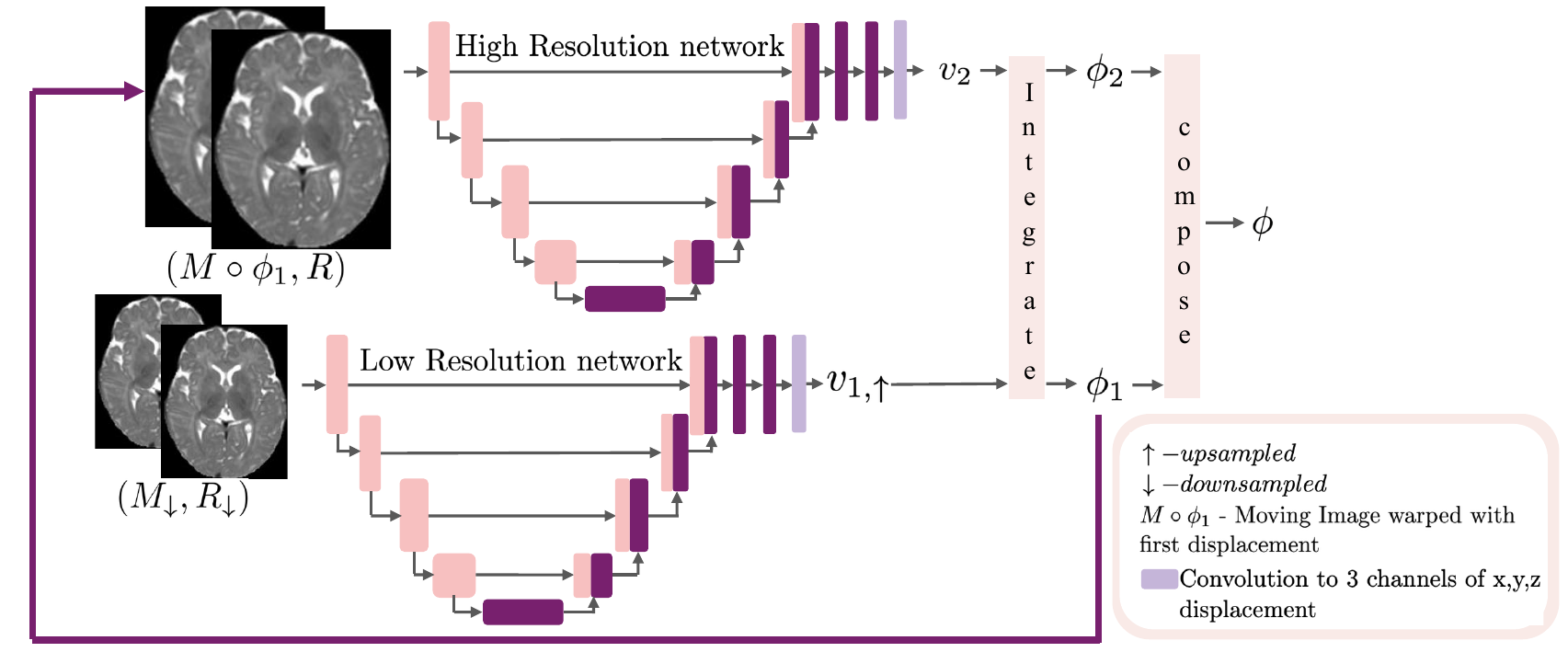}
\caption{Network architecture: Twin-U-Nets are trained to output stationary velocity fields at different resolutions; these are then combined to form a final deformation field that aligns the input ($M$) and target ($R$). The first network takes downsampled inputs and outputs a low-resolution deformation field, that is used to warp the full-resolution moving image (as indicated by the magenta arrow). This warped image is then processed by the high-resolution network to learn the residual deformation.} 
\label{fig1}
\end{figure}

Traditional image registration methods handle large deformations – such as those required for modelling gyrification – by adopting coarse-to-fine deformation strategies. This was adapted for deep learning by ~\cite{Mok_Chung_2020}. We adopt the same approach, training twin U-Nets at two resolutions to learn a multi-resolution stationary velocity field (Figure \ref{fig1}). Diffeomorphisms are enforced through scaling and squaring layers, similarly to VoxelMorph~\cite{Dalca2019UnsupervisedSurfaces} .

The low-resolution network learns large-range deformations. It takes as input downsampled versions of the moving ($M_{\downarrow}$) and reference ($R_{\downarrow}$) volumes and outputs a stationary velocity field (SVF), $v_1$. This field is then upsampled, integrated using scaling and squaring, and applied to deform the full-resolution moving image, $M$. Following~\cite{Mok_Chung_2020}, the low-resolution network is trained independently, after which its weights are frozen.

The output of this low-resolution network ($M \circ \phi_1$) is subsequently used as the moving input to a high-resolution network, alongside the full-resolution reference image ($R$). This network handles finer details of the deformation following the coarser alignment of the low-resolution network, and outputs a second SVF, $v_2$, which encodes the residual deformation. This field is then integrated and composed with the first field.
\begin{equation}
\phi = \phi_{1} \circ \phi_2 + \phi_2
\end{equation}
The final deformation field, $\phi$, is then applied to warp the full-resolution, unmoved moving image $M$ via a single interpolation step ($M \circ \phi $). Here, $\circ$ represents warping using a spatial transformer.

\subsubsection{Optimisation}
The network is trained to minimise a multi-component loss:
\begin{equation} \mathcal{L} = \mathcal{L}_{sim} + \lambda_1\mathcal{L}_{bio}  + \lambda_2\mathcal{L}_{smooth}
\label{loss} 
\end{equation}
Here, $\mathcal{L}_{sim}$ is a data similarity term that  minimises the mean squared differences (MSE) between the intensities of $ M\circ\phi$ and $R$, for each voxel; and $\mathcal{L}_{smooth}$ is a diffusion-based regulariser which encourages smooth deformations by penalising the spatial gradients of the displacement $\mathbf{u} = \phi - \textbf{I}d$, over $N$ voxels as:
\begin{equation} 
\mathcal{L}_{smooth} = \frac{1}{N}\sum\|\nabla\textbf{u}\|^2 \label{loss:diffusionregulariser} 
\end{equation}
$\mathcal{L}_{bio}$ is a biomechanical loss term based on the Neo-Hookean model for hyperelasticity. It considers the deformation in terms of the Jacobian of the deformation map $\phi$, denoted as $\mathbf{F}$. Following \cite{Rodriguez_Hoger_McCulloch_1994}, growth is induced by decomposing $\mathbf{F}$ into $\mathbf{F_g}$, the prescribed growth and $\mathbf{F_e}$, the elastic response needed to maintain mechanical equilibrium after the imposed remodelling:
\begin{equation}
\mathbf{F} = \mathbf{F_e}  \mathbf{F_g}
\label{equation:decomposition} 
\end{equation}
Previous work \cite{DaSilva2021DistinguishingNetworks} derived a differentiable implementation of this model, creating a loss that minimises the strain energy density of the deformation. 
\begin{equation}
\mathcal{L}_{bio} = \sum \left[ \frac{\mu}{2} \left( \text{Tr} \left( \mathbf{F_e} \mathbf{F_e}^{\textrm{T}} \right) J_e^{-\frac{2}{3}} - 3 \right) + \frac{\kappa}{2} (J_e-1)^2 \right]
\label{equation:StrainEnergyloss} 
\end{equation}
Here, $\mu$ is the shear modulus (assigned as 0 for the background, 0.01 for CSF, and 1 for brain tissue), and $\kappa$ is the bulk modulus, set as $\kappa = 100\mu$ in~\cite{da_silva_biomechanical_2020,DaSilva2021DistinguishingNetworks}. $J_e = \det(\mathbf{F_e})$ denotes the Jacobian determinant of the elastic response. Subject-specific growth maps are prescribed isotopically as $\mathbf{F_g} = (g^{-1/3})\textbf{I}$, where $g= V_{t1,\text{seg}} / V_{t2,\text{seg}}$ is estimated from the relative volume change from time point 1 to time point 2, for each of the segmented structures (cortical grey matter, white matter, ventricles, cerebellum, deep grey matter, brainstem and hippocampi \& amygdala), after linear alignment. The computed value is assigned uniformly to all voxels within the corresponding segmented structure.

\subsubsection{Training and implementation}
64 subjects were used for training, 9 for hyper-parameter tuning (validation) and 19 were left out for testing. The framework was implemented in PyTorch \cite{paszke2019pytorch}. Each network was trained for 1500 epochs and optimised through an  AdamW optimiser \cite{loshchilovdecoupled2019} with a learning rate of $1 \times 10^{-5}$.  We set $\lambda_1 = 0.01$ and $\lambda_2 = 10^{-5}$.

\subsubsection{Evaluation methods}

Model alignment accuracy was assessed using Dice score overlap between deformed tissue segmentations and the reference follow-up segmentations. Negative Jacobian determinants were quantified (within brain tissues) to assess the proportion of topology-breaking transformations. For comparison, we selected two state-of-the-art registration baselines: one learning-based (VoxelMorph \cite{balakrishnan_voxelmorph_2019}) and one classical (Elastic SyN from ANTsPy \cite{avants2009advanced}). To evaluate adherence to real growth, we compared cortical grey matter and white matter volumes to population-level volume–age trends. We also compared predicted volumes to ground truth follow-up volumes by computing the Absolute Symmetric Percentage Volume Change (ASPVC), which indicates minimal difference when closer to zero.
\begin{equation}
ASPVC = 100 \frac{|V_2 - V_1|}{0.5 (V_1 + V_2)}
\end{equation}
Only relative growth was considered, as absolute volume changes are suppressed by linear registration.

\subsubsection{Ablation Study}
The performance of each of the network components was evaluated through an ablation study. The model was tested with and without the biomechanical loss. The chosen network architecture was compared to two alternative setups: one using a single network, and another using two networks trained simultaneously, instead of pre-training the low-resolution network.

\section{Results and Discussion}

\subsubsection{Alignment Accuracy} 

\begin{figure}
\includegraphics[width=\textwidth]{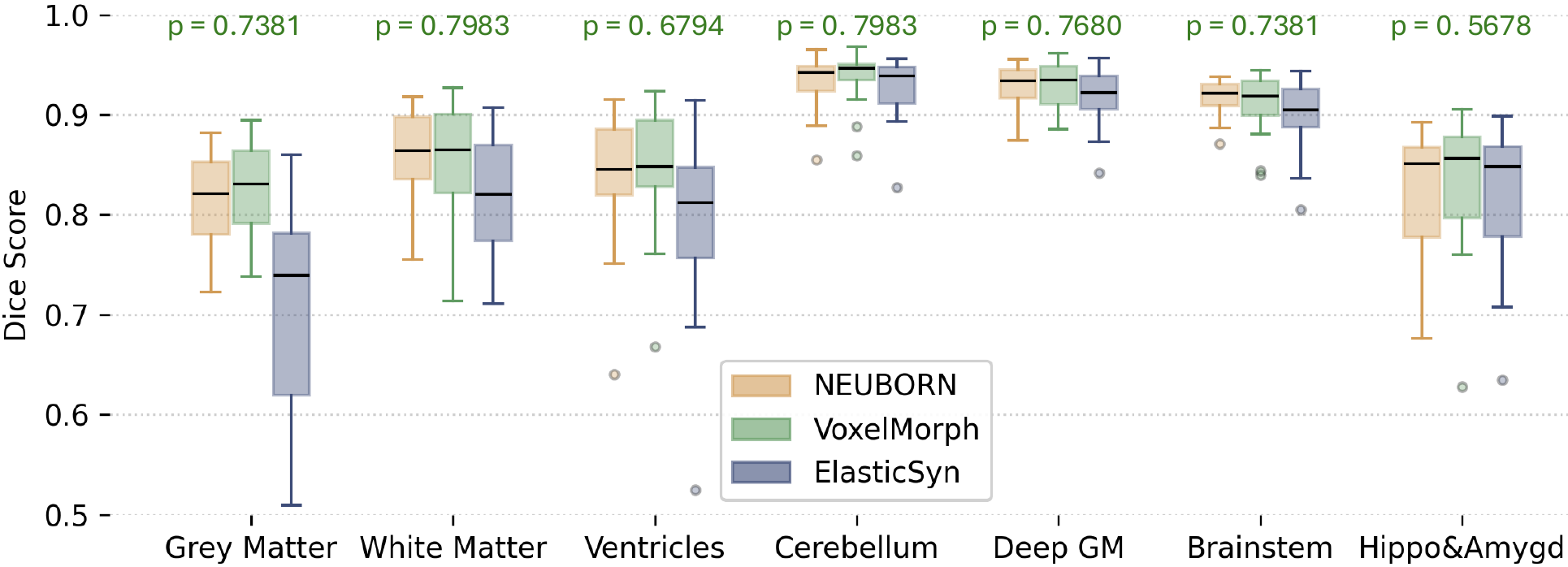}
\caption{Boxplots comparing dice scores for each tissue segmentation per model. We include the per tissue p-value of paired t-tests comparing the performance of NEUBORN and VoxelMorph.} 
\label{fig2}
\end{figure}

Table \ref{tab:registration_metrics} presents a summary of quantitative results and Figure \ref{fig2} shows a boxplot comparing the performance of each model per tissue. Our model (Dice=87.3) performs comparably to VoxelMorph (Dice=87.8) in terms of alignment accuracy with paired t-tests indicating no statistically significant difference in performance in any tissue (average p = 0.7269) (See Fig \ref{fig2}). While alignment is comparable, NEUBORN produces significantly fewer negative Jacobians (approximately four orders of magnitude fewer than VoxelMorph). Negative Jacobians in the deformation maps indicate folding or tearing of the tissue, which is anatomically impossible, therefore mappings that are more diffeomorphic, naturally translate to being anatomically more realistic. By contrast, all ElasticSyn transformations are fully diffeomorphic; however, registration accuracy falls considerably behind; even when reducing elastic regularisation, similar dice alignment was achieved.

\begin{table}
\centering
\caption{Comparison of registration models. The first column shows the average Dice overlap per tissue. The second column reports the average number of non-background negative Jacobians found per image. The third column presents the average percentage of non-background negative Jacobians. The last column shows the ASPVC of the cortical volume. For each metric, the standard deviation is reported in parentheses.}
\label{tab:registration_metrics}
\begin{tabular}{|c|c|c|c|c|}
\hline
\textbf{Model} & \textbf{Dice (\%)} & $\mathbf{J_{\phi}} \le 0$ & $\mathbf{J_{\phi}} \le 0\ \text{(\%)}$ & $ASPVC_{\text{cortex}}$ \\
\hline

NEUBORN        & 87.3(0.0016) & 0.07(0.06)  & $3.47\times 10^{-6}(3.25 \times 10^{-6})$ & 3.82(0.24)\\
VoxelMorph                 & 87.8(0.0004)  & 853.5(43.2)  & 0.04(0.002) & 7.73(0.27)\\
ElasticSyn         & 84.0 (0.0000) & 0.0(0.0) & 0.0(0.0) & 21.7(0.01)\\
\hline
\end{tabular}
\end{table}

\begin{figure}[h]
\includegraphics[width=\textwidth]{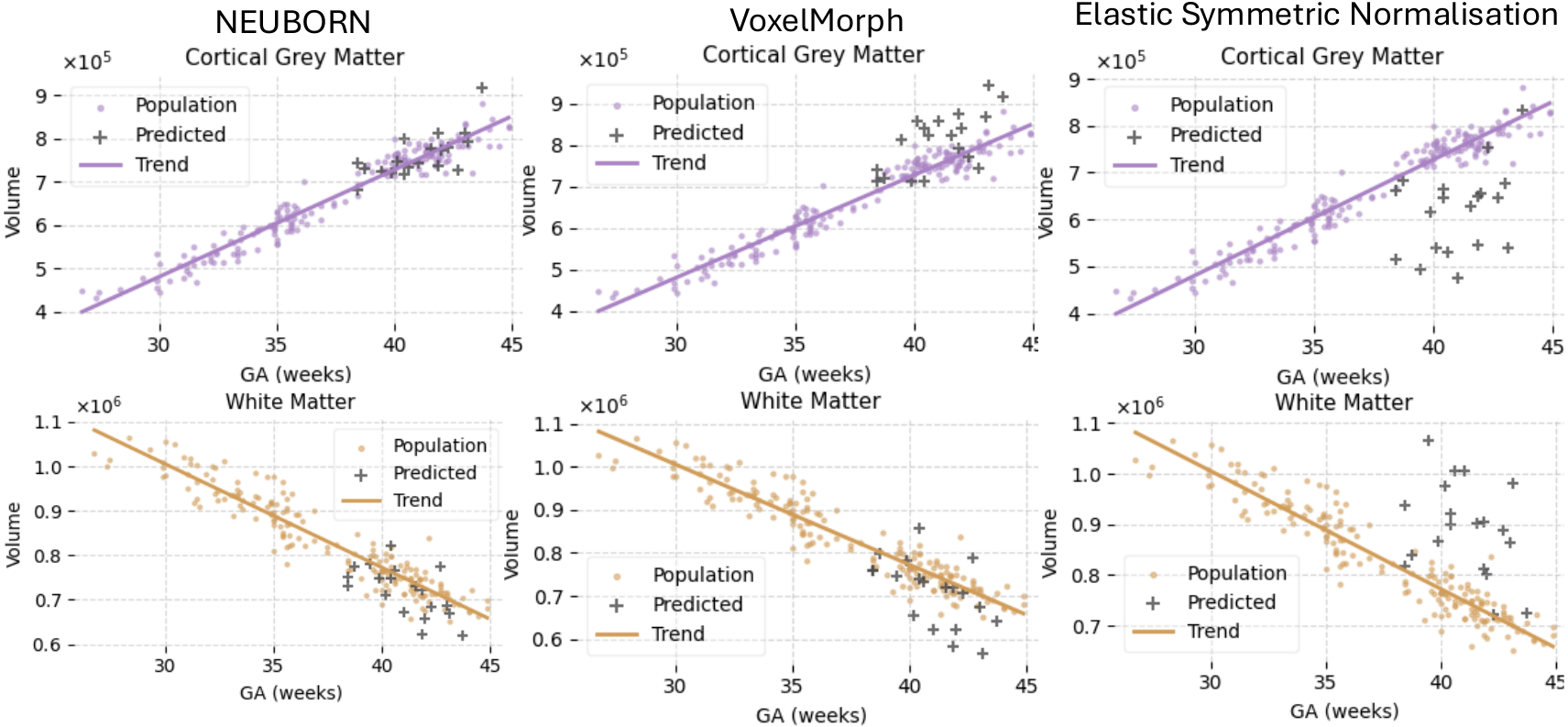}
\caption{Model predictions (in grey) plotted against population trends computed from all 92 available subjects (purple = grey matter, yellow = white matter) for NEUBORN, VoxelMorph, and ElasticSyn.}
\label{fig4}
\end{figure}

Figure~\ref{fig4} shows the age–volume relationship observed in the data as the brains grow, for both grey and white matter volumes. The simulated follow-up volumes of the 19 left-out test subjects, computed by each model, are displayed in grey. NEUBORN’s predictions align with the age–volume trend, demonstrating that NEUBORN generates deformation fields that grow brains in a way that faithfully follows population-level growth trajectories. When comparing the simulated follow-up volumes directly to the ground truth, the $ASPVC_{\text{cortex}}$ computed from NEUBORN's volumes is 3.82 — close to 0 — indicating that individual volumes are grown very closely in line with their true individual trajectories. In this, the model demonstrates that it can generalise well to new examples, even at the subject-specific level, and is robust to cortical shape heterogeneity, even within this small group.

By contrast, the volumes of VoxelMorph’s simulated follow-ups deviate from the population-level trend and show a larger difference from the ground truth volumes compared to NEUBORN ($ASPVC_{\text{cortex}} = 7.73$). VoxelMorph’s predictions appear biased towards overestimating cortical volume at the expense of underestimating white matter. A Wilcoxon signed-rank test showed no statistically significant difference between the ground truth and NEUBORN cortical volumes ($p = 0.05$), but did find a significant difference for VoxelMorph ($p = 0.0046$). This tendency to overestimate cortical volume could explain its marginal improvement in Dice overlap. However, both methods show statistically significant differences in white matter volume. ElasticSyn follows the opposite trend to VoxelMorph and generally displays much larger volume differences from the ground truth than the other methods ($ASPVC_{\text{cortex}} = 21.7$).

\subsection{Ablation Study}

Table~\ref{tab:ablation} shows that NEUBORN achieves the highest Dice score (87.3\%) and adheres closely to true anatomical trajectories (low ASPVC\textsubscript{cortex}), with negligible folding. Removing the biomechanical loss leads to increased folding artifacts and deviation from developmental trends. Training the two networks simultaneously, as well as using only one network at the highest resolution moderately degrades performance across all metrics, confirming the value of the hierarchical approach.
\begin{table}
\centering
\caption{Ablation study evaluating the impact of removing the biomechanical loss, training both networks simultaneously and using only a single network.}
\label{tab:ablation}
\begin{tabular}{|c|c|c|c|c|}
\hline
\textbf{Model} & \textbf{Dice (\%)} & $\mathbf{J_{\phi}} \le 0$ & $\mathbf{J_{\phi}} \le 0\ \text{(\%)}$ & $ASPVC_{\text{cortex}}$ \\
\hline

\textbf{NEUBORN}        & \textbf{87.3(0.002)} & \textbf{0.07(0.06)}  & \textbf{3.47$\cdot$10\textsuperscript{-6}(3.25$\cdot$10\textsuperscript{-6})} & 3.82(0.24)\\
No biomechanical loss    & 87.0(0.002) & 22.0(8.61) & 1.09$\cdot$10\textsuperscript{-3}(4.25$\cdot$10\textsuperscript{-4}) & 16.8(0.78)\\
Training networks jointly     & 86.8(0.002) & 0.72(1.00) & 3.60$\cdot$10\textsuperscript{-5} (4.94$\cdot$10\textsuperscript{-5}) & 6.04(0.69)\\
Using a single network            & 85.2(0.002) & 0.0(0.0) & 0.0(0.0) & 4.17(0.34)\\
\hline
\end{tabular}
\end{table}

\section{Conclusions and Future work}

This work presents NEUBORN, a novel biomechanically constrained deep learning framework for longitudinal image registration. By integrating a diffeomorphic registration strategy with a physically grounded hyperelastic loss, the model generates anatomically plausible deformations for the full brain that accurately preserves complex subject-specific cortical growth trajectories.

Our method matches the alignment accuracy of state-of-the-art registration techniques while dramatically reducing anatomically implausible, non-diffeomorphic mappings. By incorporating the tissue’s mechanical properties during development, NEUBORN produces deformations that remain consistent with both individual cortical morphology and broader population-level growth trends. The model generalises well to new subjects, even under a reduced training regime, and improves the biological interpretability of predicted trajectories.

Accurately recovering subject-specific gyrification patterns is particularly valuable in contexts where subtle deviations from typical development may indicate early signs of neurological disorder. Future work will focus on leveraging the biomechanical interpretability to model intermediate developmental states between scans, and to enable forward projections of future brain changes at the patient level.

\begin{credits}

\subsubsection{\ackname}  The dHCP neonatal dataset was provided by the developing Human Connectome Project, KCL-ImperialOxford Consortium funded by the European Research Council (ERC) under the European Union Seventh Framework Programme (FP/2007-2013) / ERC Grant Agreement no. [319456]. We are grateful to the families who generously supported this trial. The authors acknowledge use of the King's Computational Research, Engineering and Technology Environment (CREATE) (https://doi.org/10.18742/rnvf-m076). ECR acknowledges philanthropic support from Heart of Racing LLC to the Brain Health in Gen2020 programme at King’s College London and support from the Medical Research Council (UKRI534). JDT is supported by core funding from the Wellcome/EPSRC Centre for Medical Engineering [WT203148/Z/16/Z] and by the National Institute for Health and Care Research (NIHR) Clinical Research Facility (CRF) and HealthTech Research Centre in Cardiovascular and Respiratory Medicine (HRC) at Guy’s and St Thomas’ NHS Foundation Trust. The views expressed are those of the author(s) and not necessarily those of the NHS, the NIHR or the Department of Health and Social Care. NB, MDS and AS would like to acknowledge funding from the EPSRC Centre for Doctoral Training in Smart Medical Imaging (EP/S022104/1).

\end{credits}
%


\bibliographystyle{splncs04}
\bibliography{references}
\end{document}